\documentclass[%
 reprint,
 amsmath,amssymb,
 aps,
prl,
]{revtex4-2}

\usepackage{graphicx}
\usepackage{dcolumn}
\usepackage{bm}
\usepackage{textcomp}

\begin{document}

\preprint{APS/123-QED}

\title{Phase Space Reconstruction from Accelerator Beam Measurements Using Neural Networks and Differentiable Simulations}

\author{R. Roussel}
\email{rroussel@slac.stanford.edu}
\author{A. Edelen}
\author{C. Mayes}
\author{D. Ratner}
\affiliation{SLAC National Accelerator Laboratory, Menlo Park, CA 94025, USA}
\author{J.P. Gonzalez-Aguilera}
\affiliation{Department of Physics, University of Chicago, Chicago, Illinois 60637, USA}
\author{S. Kim, E. Wisniewski, J. Power}
\affiliation{Argonne National Laboratory, Argonne, Illinois 60439, USA}

\date{\today}

\begin{abstract}
Characterizing the phase space distribution of particle beams in accelerators is a central part of accelerator understanding and performance optimization. However, conventional reconstruction-based techniques either use simplifying assumptions or require specialized diagnostics to infer high-dimensional ($>$ 2D) beam properties. In this Letter, we introduce a general-purpose algorithm that combines neural networks with differentiable particle tracking to efficiently reconstruct high-dimensional phase space distributions without using specialized beam diagnostics or beam manipulations. We demonstrate that our algorithm accurately reconstructs detailed 4D phase space distributions with corresponding confidence intervals in both simulation and experiment using a single focusing quadrupole and diagnostic screen. This technique allows for the measurement of multiple correlated phase spaces simultaneously, which will enable simplified 6D phase space distribution reconstructions in the future. 
\end{abstract}

\maketitle

Increasingly precise control of the distribution of particles in position-momentum phase space is needed for emerging applications of accelerators \cite{nagaitsev_accelerator_2021}. This includes, for example, new experiments at free electron lasers \cite{emma_first_2010,Haoyuan_generation_2021,Sun_realizing_2020,Marinelli_high_2015, Decker_tunable_2022} and novel acceleration schemes that promise higher-energy beams in compact spaces \cite{noauthor_advanced_2016}. 
Numerous techniques have been developed for precision shaping of beam distributions \cite{ha2022bunch}; however, the effectiveness of these techniques relies on accurate measurements of the 6D phase space distribution, which is a challenging task unto itself.

Tomographic measurement techniques are used in accelerators to determine the density distribution of beam particles in phase space $\rho(x, p_x, y, p_y, z, p_z)$ from limited measurements \cite{McKee_phase_1995,Hancock_tomographic_1999,Stratakis_phase_2007,yakimenko_electron_2003,rohrs_time-resolved_2009, gordon2022four}. 
The simplest form of this uses scalar metrics, such as second-order moments, to describe observations of the transverse beam distribution when projected onto a scintillating screen. \cite{Green_implementation_2015,prat_four-dimensional_2014,Mostacci_chromatic_2012}.
This process however discards significant amounts of information about the beam distribution captured by high-resolution diagnostic screens and only predicts scalar quantities of the beam distribution.
In contrast, methods using projections of the beam image, including filtered back-projection \cite{webb_physics_1987, yakimenko_electron_2003}, algebraic reconstruction \cite{kak_principles_2001,wolski2020transverse, wolski2022transverse}, and maximum entropy tomography (MENT) \cite{Hock_A_2013,rohrs_time-resolved_2009} produce more accurate reconstructions.

The MENT algorithm is particularly well-suited to reconstructing beams from limited and/or partial information sources about the beam distribution, as is the case in most experimental accelerator measurements.
MENT solves for a phase space distribution that maximizes entropy (and, as a result, likelihood), subject to the constraint that the distribution accurately reproduces experimental measurements.
While these techniques have been shown to effectively reconstruct 2D phase spaces from image projections using algebraic methods, application to higher-dimensional spaces requires independence assumptions between the phase spaces of principal coordinate axes $(x, y, z)$, complicated phase space rotation procedures \cite{hock2013tomographic, wolski2020transverse}, or simultaneous measurement of multiple 2D sub-spaces with specialized diagnostic hardware \cite{wong_4d_2022}.

Numerical optimization methods can also be used to infer beam distributions from experimental data.
For example, arbitrary beam distributions can be parameterized by a set of principal components \cite{scheinker_adaptive_2021} whose relative weights can be optimized to produce a beam distribution that, when tracked through a simulation, reproduces experimental measurements.
Alternatively, heuristics can be used to delete or generate particles in a distribution until particle tracking results match experiments \cite{wang_four-dimensional_2019, hermann_electron_2021}.
Unfortunately, these methods suffer from increasing computational cost when extending them to reconstructing high-dimensional phase space distributions, primarily due to the cost associated with optimizing the large number of free parameters needed to represent detailed beam characteristics in high-dimensional phase spaces. 

\begin{figure*}[ht]
    \includegraphics[width=\linewidth]{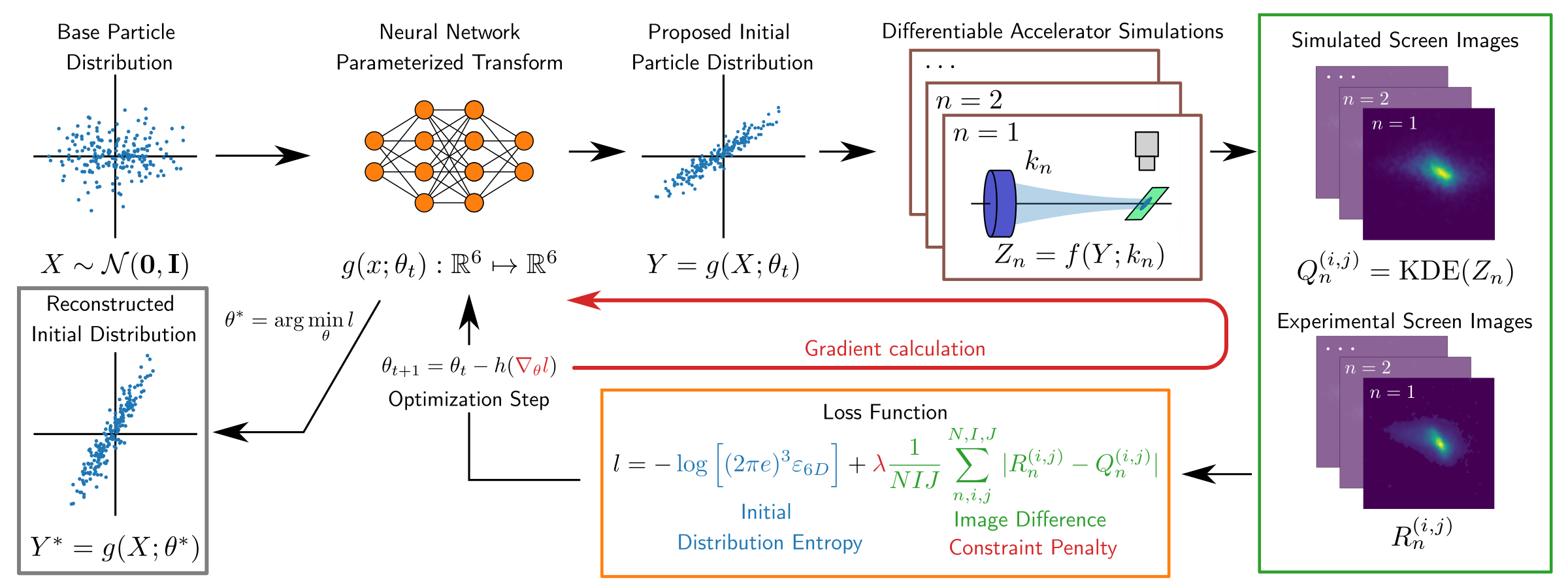}
    \caption{Description of our approach for reconstructing phase space beam distributions.
    First, a 6D base distribution is transformed via neural network, parameterized by $\theta_t$, into a proposed initial distribution. This distribution is then transported through a differentiable accelerator simulation of the tomographic beamline. The quadrupole is scanned to produce a series of images on the screen, both in simulation and on the operating accelerator. The images produced both from the simulation $Q^{(i,j)}_n$ and the accelerator $R^{(i,j)}_n$ are then compared with a custom loss function, which attempts to maximize the entropy of the proposal distribution, constrained on accurately reproducing experimental measurements. This loss function is then used to update the neural network parameters $\theta_{t} \to \theta_{t+1}$ via gradient descent. The neural network transformation that minimizes the loss function generates the beam distribution that has the highest likelihood of matching the real initial beam distribution. }
   \label{fig:cartoon}
\end{figure*}

In this Letter we describe a new method that provides detailed reconstructions of the beam phase space using simple and widely-available accelerator elements and diagnostics. 
To achieve this, we take advantage of recent developments in machine learning to introduce two new concepts (shown in Fig.~\ref{fig:cartoon}): a method for parameterizing arbitrary beam distributions in 6D phase space, and a differentiable particle tracking simulation that allows us to learn the beam distribution from arbitrary downstream accelerator measurements.
We examine how this method extracts detailed 4-dimensional phase space distributions from measurements in simulation and experiment, using a simple diagnostic beamline, containing a single quadrupole, drift and diagnostic screen to image the transverse ($x,y$) beam distribution.  
Finally, we discuss current limitations of this method as well as future directions for the design of novel accelerator diagnostics using this technique.

\begin{figure*}[ht]
    \includegraphics[width=\linewidth]{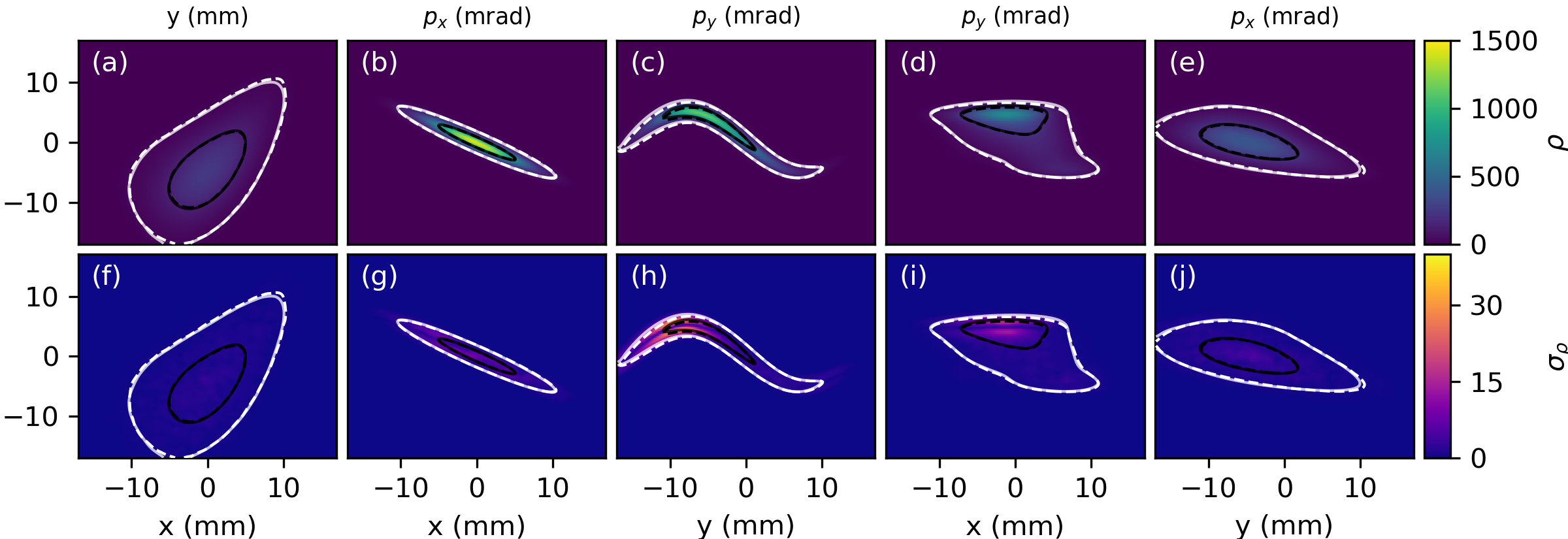}
    \caption{Comparisons between the synthetic and reconstructed beam probability distributions using our method. (a-e) Plots of the mean predicted phase space density projections in 4D transverse phase space. Contours that denote the 50$^{th}$ (black) and 95$^{th}$ (white) percentiles of the synthetic ground truth (dashed) and reconstructed (solid) distributions. (f-j) Plots of the predicted phase space density uncertainty.}
   \label{fig:synthetic_reconstruction}
\end{figure*}

We first demonstrate our algorithm using a synthetic example, where we attempt to determine the distribution of a 10-MeV beam given a predefined structure in 6D phase space.
The propagation of a synthetic beam distribution through a simple diagnostic beamline containing a 10 cm long quadrupole followed by a 1.0 m drift is simulated using a custom implementation of Bmad \cite{Sagan:Bmad2006} referred to here as Bmad-X.
To illustrate the capabilities of our technique, the synthetic beam contains multiple higher order moments between each phase space coordinates (see Supplemental Materials for details). 
To simulate an experimental measurement, we simulate particles traveling through the diagnostic beamline while the quadrupole strength $k$ is scanned over $N$ points.
The final transverse distribution of the beam is measured at each quadrupole strength using a simulated $200 \times 200$ pixel screen, with a pixel resolution of 300 \textmu m (image data can be viewed in the Supplemental Materials).
The set of images, where the intensity of pixel ($i,j$) on the n'th image is represented by $R^{(i,j)}_n$, is then collected with the corresponding quadrupole strengths to create the data set, which is then split into training and testing subsets by selecting every other sample as a test sample, resulting in 10 samples for each data subset.

The reconstruction algorithm begins with the generation of arbitrary initial beam distributions (referred to here as proposal distributions) through the use of a neural network transformation.
A neural network, consisting of only 2 fully-connected layers of 20 neurons each, is used to transform samples drawn from a 6D normal distribution centered at the origin to macro-particle coordinates in real 6D phase space (where positional coordinates are given in meters and momentum coordinates are in radians for transverse momenta).
As a result, the coordinates of particles in the proposal distribution are fully parameterized by the neural network parameter set $\theta_t$.

Fitting neural network parameters to experimental measurements is done by minimizing a loss function to determine the most likely initial beam distribution, subject to the constraint that it reproduces experimental measurements; this is similar to the MENT algorithm \cite{Hock_A_2013}.
The likelihood of an initial beam distribution in phase space is maximized by maximizing the distribution entropy, which is proportional to the log of the 6D beam emittance $\varepsilon_{6D}$ \cite{lawson1973emittance}.
Thus, we specify a loss function that minimizes the negative entropy of the proposal beam distribution, penalized by the degree to which the proposal distribution reproduces measurements of the transverse beam distribution at the screen location.
To evaluate the penalty for a given proposal distribution, we track the proposal distribution through a batch of accelerator simulations that mimic experimental conditions to generate a set of simulated images $Q^{(i,j)}_n$ to compare with experimental measurements. 
The total loss function is given by
\begin{equation}
l= -\log \Big[(2\pi e)^3\varepsilon_{6D}\Big] + \lambda\frac{1}{NIJ}\sum_{n,i,j}^{N,I,J}|R^{(i,j)}_n - Q^{(i,j)}_n|
\label{eqn:loss_fn}
\end{equation}
where $\lambda$ scales the distribution loss penalty function relative to the entropy term and is chosen empirically based on the resolution of the images. 

However, the large ($> 10^3$) number of free parameters contained in the neural network transformation used to generate proposal distributions necessitates the use of gradient-based optimization algorithms such as Adam \cite{kingma_adam_2017} to minimize the loss function.
Thus, we need to implement computation of the loss function such that it supports backward differentiation \cite{lecun_efficient_2012} (referred to here as \textit{differentiable} computations), allowing us to cheaply compute loss function derivatives with respect to every neural network parameter.
This requires that every step involved in calculating the loss function is also differentiable, including computing the beam emittance and tracking particles through the accelerator.
Unfortunately, to the best of our knowledge, no particle tracking codes currently support backwards differentiation.
To satisfy this requirement, we implement particle tracking in Bmad-X using the machine learning library \emph{PyTorch} \cite{paszke_pytorch_2019}.
We estimate screen pixel intensities from a discrete particle distribution with a differentiable implementation of kernel density estimation \cite{rosenblatt1956remarks}. 

\begin{figure}[ht]
    \includegraphics[width=\linewidth]{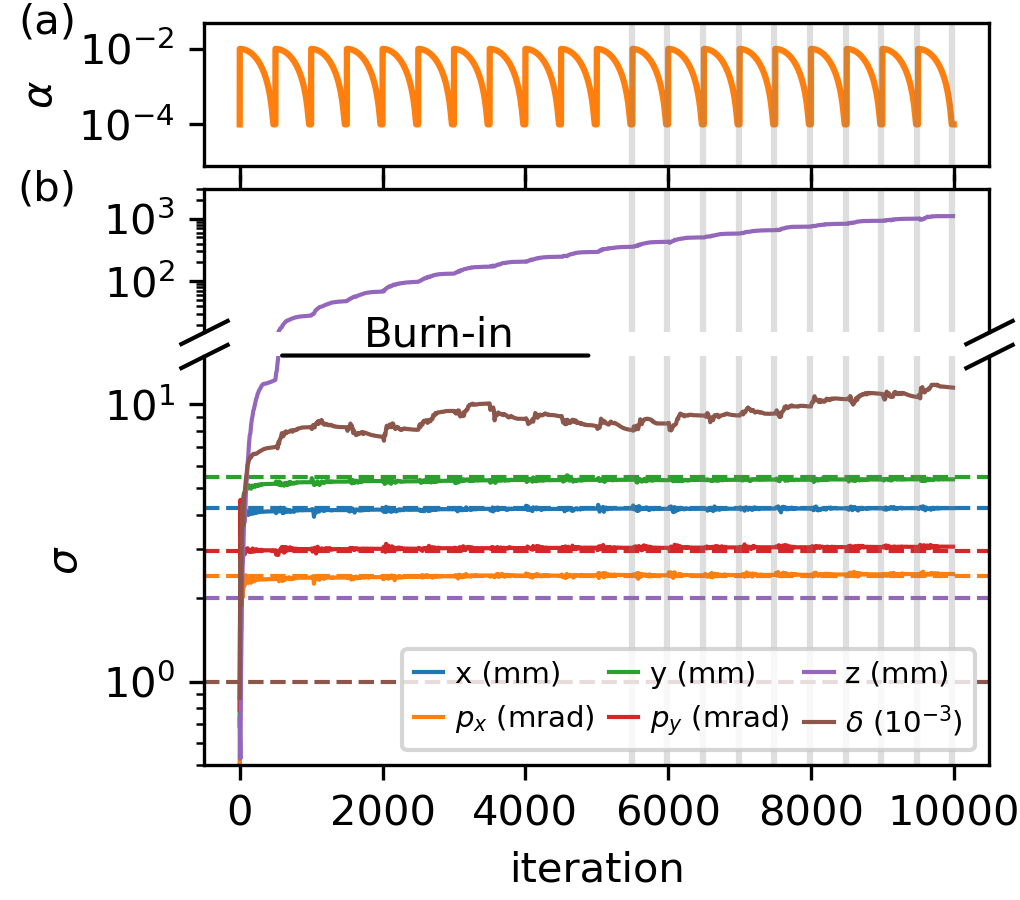}
    \caption{Evolution of the proposal distribution during training on synthetic data. (a) Learning rate schedule for snapshot ensembling. 
    (b) Second order moments of beam reconstruction during training for each phase space coordinate. Dashed lines denote ground truth values. Vertical lines denote snapshot locations after burn-in period. 
    }
   \label{fig:convergence}
\end{figure}

Results from our reconstruction of the initial beam phase space using synthetic images are shown in Fig.~\ref{fig:synthetic_reconstruction}.
We characterize the uncertainty of our reconstruction using snapshot ensembling \cite{huang2017snapshot}.
During model training, we cycle the learning rate of gradient descent in a periodic fashion which encourages the optimizer to explore multiple possible solutions (if they exist).
After several of these cycles (known as a ``burn-in" period), we save model parameters at each minima of the learning rate cycle, as shown in Fig.~\ref{fig:convergence}(a).
We then weight predictions from each model equally, using them to predict a mean initial beam density distribution Fig.~\ref{fig:synthetic_reconstruction}(a-e) with associated confidence intervals Fig.~\ref{fig:synthetic_reconstruction}(f-j).
Performing this analysis by tracking $10^5$ particles for each image took less than 30 seconds per ensemble sample using a professional grade GPU ($<60$ ms per iteration, 500 steps per ensemble sample).

\begin {table}[htp]
\caption{Predicted Emittances Compared to True Values}
\label{TB:synth}
\begin{ruledtabular}
\begin {tabular}{l l l l l}
Parameter & \begin{tabular}{@{}l@{}} Ground\\truth \end{tabular} & \begin{tabular}{@{}l@{}} RMS\\Prediction \end{tabular} & Reconstruction & Unit \\
\colrule
$\varepsilon_{x}$ & 2.00 & $2.47$ & $2.00 \pm 0.01$ & mm-mrad\\
$\varepsilon_{y}$ & 11.45 & $14.10$ & $10.84\pm 0.04$ & mm-mrad\\
$\varepsilon_{4D}$ & 18.51 & $34.83$$^*$ & $17.34 \pm 0.08$ & mm$^2$-mrad$^2$\\
\end {tabular}
\end{ruledtabular}
$^*$ Assumes x-y phase space independence
\end{table}  

We see excellent agreement between the average reconstructed and synthetic projections in both transverse correlated and uncorrelated phase spaces.
Furthermore, the prediction uncertainty from ensembling is on the order of a few percent relative to the predicted mean, providing confidence that the overall solution found during optimization is unique.
As shown in Table \ref{TB:synth}, reconstructions of the beam distribution from image data predicts transverse phase space emittances that are closer to ground truth values than those predicted from second-order moment measurements of the transverse beam distribution.
This results from non-linearities and cross-correlations present in the 4-D transverse phase space distribution.

It is instructive to examine the evolution of the proposal distribution during model training.
In Fig.~\ref{fig:convergence}(b) we examine second order scalar metrics of the proposal distribution after each training iteration for each phase space coordinate.
The entropy term in Eq.~\ref{eqn:loss_fn} causes the distribution to expand in 6D phase space until constrained by experimental evidence.
Phase space components that have the strongest impact on beam transport through the beamline as a function of quadrupole strength converge quickly to the true values, whereas the ones that have little-to-no impact (e.g. the longitudinal distribution characteristics) continue to grow.
In other cases, there is weak coupling between the experimental measurements and beam properties; for example, chromatic focusing effects due to the energy spread $\sigma_\delta$ of the beam weakly affect the measured images.
Here, the reconstruction can only provide an upper-bound estimate of the energy spread, since small changes in transverse beam propagation due to chromatic aberrations are overshadowed by statistically dominated particle motion.
Convergence of the proposal distribution thus provides a useful indicator of which phase space components can be reliably reconstructed from arbitrary sets of measurements.

We now describe a demonstration of our method on an experimental example at the Argonne Wakefield Accelerator (AWA) \cite{conde2017research} facility at Argonne National Laboratory.
Our objective is to identify the phase space distribution of 65-MeV electron beams at the end of the primary accelerator beamline.
The focusing strength of a quadrupole, with an effective length of 12 cm, is scanned while imaging the beam at a transverse scintillating screen located 3.38 m downstream.
Charge windowing, image filtering, thresholding and downsampling were used to generate a set of 3 images for each quadrupole setting (see the Supplemental Materials for additional details).

\begin {table}[htp]
\caption{Predicted Emittances from Experimental Data}
\label{TB:exp}
\begin{ruledtabular}
\begin {tabular}{l l l l}
Parameter & \begin{tabular}{@{}l@{}} RMS\\Prediction \end{tabular} & Reconstruction & Unit \\
\colrule
$\varepsilon_{x,n}$ & $4.18 \pm 0.71$ & $4.23 \pm 0.02$ & mm-mrad\\
$\varepsilon_{y,n}$ & $3.65 \pm 0.36$ & $3.42\pm 0.02$ & mm-mrad\\
\end {tabular}
\end{ruledtabular}
\end{table}  

\begin{figure*}[ht]
    \includegraphics[width=\linewidth]{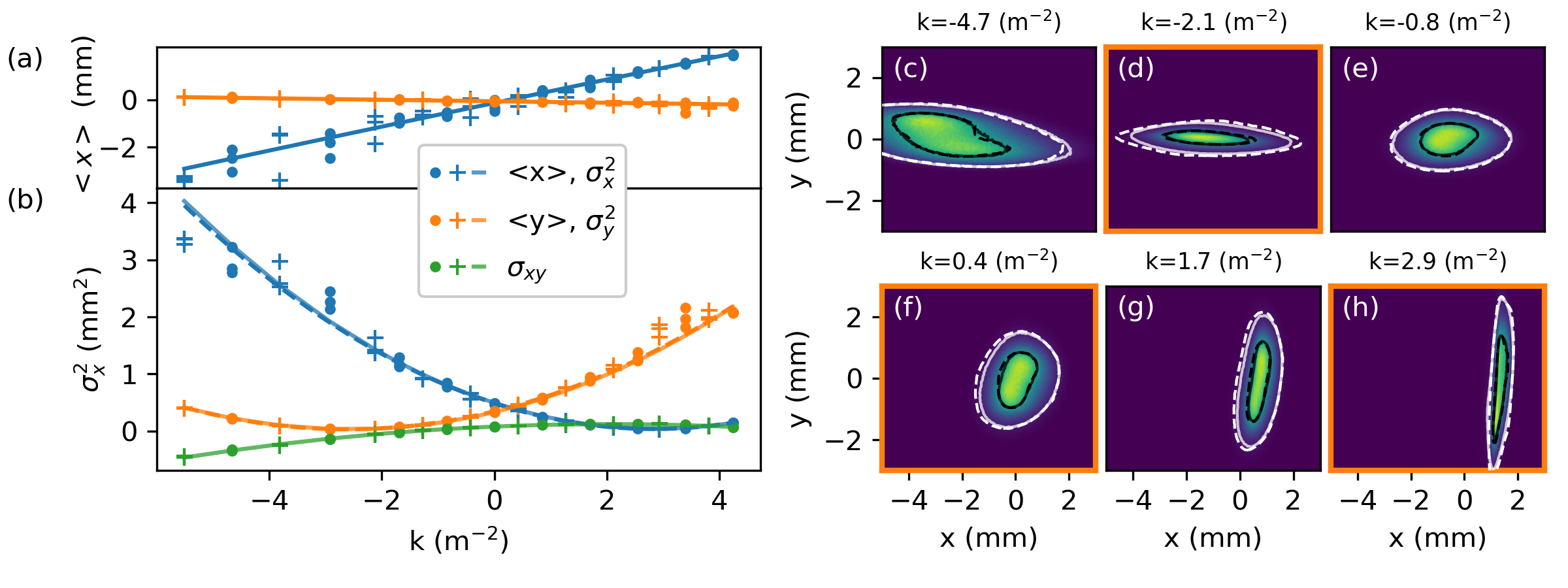}
    \caption{Reconstruction results from experimental measurements at AWA. Comparison between measured and predicted beam centroids (a) and second-order beam moments (b) on the diagnostic screen as a function of geometric quadrupole focusing strength ($k$). Points denote training samples and crosses denote test samples. Dashed line shows second order polynomial fit of training data and solid line shows predictions from image-based phase space reconstruction. We also compare (c-h) screen images and reconstructed predictions for a subset of quadrupole strengths. Contours denote the 50$^{th}$ (black) and 95$^{th}$ (white) percentiles of the measured (dashed) and predicted (solid) screen distributions. Orange borders denote test samples.}
   \label{fig:experimental_results}
\end{figure*}

We developed a differentiable simulation in Bmad-X of the experimental beamline, including  details of the diagnostics used, such as the location and properties of beamline elements and the per-pixel resolution of the imaging screen. 
With this simulation, we used our method to reconstruct the beam distribution from experimentally-measured transverse beam images. 
The results, as shown in Figure~\ref{fig:experimental_results} and Table~\ref{TB:exp}, demonstrate good agreement between experimental measurements of the beam distribution and predictions from our reconstruction. 
Scalar predictions of the beam emittances from the image-based reconstruction are consistent with those calculated from RMS measurements. 
Additionally, our reconstruction method accurately reproduces fine features of the transverse beam distribution that were not present in the training data set.

In this work, we have demonstrated how differentiable particle tracking simulations, combined with neural network based representations of beam distributions, can be used to interpret common image-based diagnostic measurements.
Our method produces detailed reconstructions of 4-dimensional transverse phase space distributions from limited data sets, without the use of complex phase space manipulations or specialized diagnostics.
Additionally, our reconstruction identifies limitations in resolving certain aspects of the beam distribution based on available measurements.
This analysis is enabled by inexpensive gradient calculations provided by backwards differentiable physics simulations.
As a result, we are able to determine thousands of free parameters used to describe complex beam distributions on a time scale similar to the time it takes to perform the physical tomographic measurements themselves.
Thus, our reconstruction technique is suitable for inferring detailed beam distributions in an online fashion, i.e. during accelerator operations.

As with any new algorithmic technique, there are areas for future improvement.
Uncertainty estimates provided by the reconstruction algorithm only capture systematic uncertainties from optimizing the loss function, Eq.~\ref{eqn:loss_fn}; thus it ignores systematic uncertainties of the physical measurement and stochastic noise inherent in real accelerators.
Future work will incorporate Bayesian analysis techniques into the reconstruction to provide calibrated uncertainty estimates to experimental measurements.
Also, while our method significantly increases the speed of high-dimensional phase space reconstructions, achieving this requires substantial amounts of memory to store the derivative information of each macro-particle at every tracking step ($\sim 4$ GB for each snapshot in the analysis performed here). 
Peak memory consumption can be reduced through the use of checkpointing \cite{dauvergne2006data} or pre-computing derivatives associated with tracking particles through the entire beamline. 
Finally, this method is limited by the availability of accurate, computationally efficient, backwards differentiable particle tracking simulations.
In order to expand the range of diagnostic measurements that can be analyzed by this technique, further investment in differentiable implementations of particle tracking simulations is needed.

This new reconstruction approach opens the door to efficient, detailed characterization of 6-dimensional phase space distributions and new types of compound diagnostic measurements. 
By adding longitudinal beam manipulations, such as transverse deflecting cavities paired with dipole spectrometers, to the beamline used here, full phase space distributions can be characterized through a series of quadrupole strength and deflecting cavity phase scans.

\begin{acknowledgments}
The authors would like to thank Lukas Heinrich and Michael Kagan for useful discussion during the early conceptual development of this work. This work was supported by the U.S. Department of Energy, under DOE Contract No. DE-AC02-76SF00515, the Office of Science, Office of Basic Energy Sciences and the Center for Bright Beams, NSF award PHY-1549132.
This research used resources of the National Energy Research Scientific Computing Center (NERSC), a U.S. Department of Energy Office of Science User Facility located at Lawrence Berkeley National Laboratory, operated under Contract No. DE-AC02-05CH11231 using NERSC award ERCAP0020725.

Author Contributions: R.R. and A.E. conceived of the idea to combine differentiable simulations with machine learning for phase space tomography. R.R. led the studies and performed the work for phase space reconstruction. A.E. and D.R. provided technical guidance and feedback. J.P.G. developed the differentiable simulation with guidance from R.R. and C.M.. S.K., E.W. and J.P. assisted with experimental studies at AWA. R.R. and A.E. wrote the manuscript. J.P.G., C.M. provided substantial edits to the manuscript. All authors provided feedback on the manuscript.

\end{acknowledgments}


%

\end{document}